\documentclass[amssymb, amsmath,aps,prb,reprint,
showpacs,showkeys]{revtex4-1}
\usepackage{graphicx}
\usepackage{dcolumn}%
\usepackage{bm}%
\usepackage{multirow}

\begin{document}

\title{On chemical bonding of Helium with hcp-Beryllium}

\author{A. S. Bakai}
\affiliation{National Science Centre Kharkiv Institute of Physics
and Technology, 61108 Kharkiv, Ukraine }
\author{A.N. Timoshevskii}
\altaffiliation{
Author to whom correspondence should be addressed. Electronic
address: tim@imag.kiev.ua}
\author{B.Z. Yanchitsky}
\affiliation{Institute of Magnetism, National Academy of Science
Ukraine, 03142 Kiev,Ukraine }

\begin{abstract}
Chemical inertness is the key property of helium determining its solubility,
distribution and accumulation kinetics in metals. Against all expectations, our
\textit{ab initio} calculations show a substantial chemical bonding between He and Be
atoms in the hcp-Be matrix when He occupies a non-symmetric position in a basal plane.
\end{abstract}



\maketitle

Helium is the most inert element of the periodic table. Having its first
electronic level ($1s$-shell) completely occupied, the electro-neutral state
of the atom is very stable. In an excited state, when one of the electrons
moves to the next energy level, He may form short-living dimers with
fluorine or chlorine: He-F and He-Cl. The very existence of these excimers
can be viewed as a proof of the degree of chemical inertia of helium.
Chemical inertia determines the helium behavior in the irradiated structural
materials of nuclear reactors, where He is formed in substantial amounts
due to nuclear fission reactions. Accumulation of He in metal alloys in the form
of the inert lattice gas, leads to their so-called ``helium swelling'' and embrittlement.

We studied the distribution of interstitial He in hexagonal
close-packed (hcp) beryllium by means of \textit{ab initio} modeling.
Employing a pseudopotential technique \cite{Giannozzi_2009JPCM}, we
obtained a surprising result: interstitial He forms a chemical
bonding with Be matrix \cite{Bakai_2011LTP}.

Further, using a highly accurate all-electron full-potential linearized
augmented plane wave (FLAPW) method, we confirmed results of pseudopotential
calculations, analysed partial DOSes and their integral quantities. We performed
analysis of chemical bonding in the Be-He system using Crystal Orbital Hamilton
Population (COHP) indicator \cite{Dronskowski_1993JPC}, implemented in SIESTA
package \cite{Soler_2002JPCM}. In this Letter the results of our investigations
are presented.

The Density Functional Theory (DFT) was used in Generalized
Gradient Approximation (GGA \cite{Perdew_1996PRL}),
as implemented in the \textit{Quantum-ESPRESSO} (QE) program package
\cite{Giannozzi_2009JPCM}. The same
GGA approximation was used for calculations using FLAPW method
(Wien2k software package \cite{Wien2k}).

\begin{figure}[!bp]
\includegraphics[width = 8.5 cm]{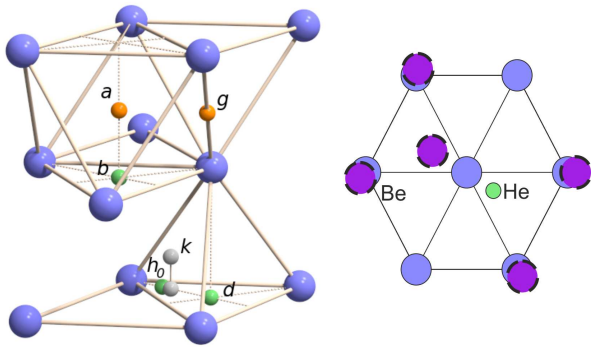}
\caption{(Color online) Equilibrium positions of He in hcp-Be and atomic
relaxation near He atom in position \textit{(h$_0$)}.
Stable positions \textit{(b,d,h$_0$)}
are shown as green circles, unstable \textit{(a,g)} as orange ones.
}
\label{FigPositions}
\end{figure}

We have chosen 22 positions of interstitial He in hcp-Be (space group 194,
two Be atoms in the unit cell at positions (1/3,2/3,1/4), (2/3,1/3,3/4)) as
starting geometry configurations for the calculations. These positions included
the centers of octahedra and tetrahedra, as well as the midpoints between the
centers of these polyhedra (Fig.~\ref{FigPositions}).
The midpoints between the two neighboring lattice
sites were also included as testing positions for interstitial He.
All positions were generated by relation:
$1/2~(p_1,p_2,p_3) + 1/2\sum_{ij}\left(\mathbf{r}_i + \mathbf{r}_j\right)$,
where summation $i,j$ is over octahedra and tetrahedra,
and $p_1,p_2,p_3$ take values $\{0,1\}$ independently.

Calculations were performed for Be$_{64}$He (4x4x2 hcp Be unit cells), 
Be$_{96}$He (4x4x3 hcp Be unit cells) 
supercells and included relaxation of atomic positions.
Atomic nuclei were moved to their equilibrium positions while forces acting on
nuclei were greater than 1~mRy/Bohr. Lattice parameters of supercells were
according to hcp-Be, that were obtained by
minimisation of the total energy with respect to $a$ and $c/a$. Calculated
lattice parameters from pseudopotential method $a=2.246$~\AA{}, $c/a=1.573$ are
comparable with experimental values $a=2.2858$~\AA{}, $c/a=1.568$
\cite{Mackay_1963JNM}, whereas FLAPW data $a=2.272$~\AA{}, $c/a=1.565$
are in adequate agreement.

On the basis of the total-energy calculations, we found five equilibrium
positions of He atom (zero force on He nucleus) in the beryllium lattice.
Fig.~\ref{FigPositions} shows these equilibrium positions as green and orange
circles. To denote positions we are using Wyckoff notations \cite{InterTables} and
fractional coordinates (Table~\ref{TabPositions}). The other 17 studied
positions of He atom in hcp-Be matrix, including the tetrahedral one, were found
to be non-equilibrium. Among 5 equilibrium positions, three positions
\textit{(b),(d),(h$_0$)} are stable (Fig.~\ref{FigPositions}a, green circles),
and two \textit{(a),(g)} are unstable (saddle points with negative curvature of
potential energy surface for He displacement along [0,0,1];
Fig.~\ref{FigPositions}a, orange circles). Both FLAPW and pseudopotentail
methods produced close values of He-Be bond length (Table~\ref{TabPositions}).
The energy of isolated helium atom, which is needed to calculate the solution
energy of He in Be lattice, was obtained by performing total energy calculations
for simple cubic lattice of He atoms with lattice
parameter $10$~\AA{}. Calculated solution energies of
interstitial He are presented in Table~\ref{TabEnergy}. The previous
\textit{ab initio} results of Ganchenkova \textit{et al.}
\cite{Ganchenkova_2009JNuM} and Zhang \textit{et al.}
\cite{Zhang_2012JPCM,Zhang_2012JNM} are also shown for comparison.

According to all studies 
\cite{Ganchenkova_2009JNuM,Zhang_2012JPCM,Zhang_2012JNM}
He solution energy is in the
range of 5.5--6.5~eV. He solution energy for octahedral cavity
(\textit{(a)} position) has the value of 6.0-6.5~eV and
is in good agreement with the value of 6.1~eV from
\cite{Zhang_2012JPCM,Zhang_2012JNM} and close to 6.8~eV reported by Cayphas
\cite{Cayphas_1997JNM} from EAM potential calculations. 
Solution energies for positions \textit{(b),(d)} are in well agreement between
researches. According to our present findings, the most energy favorable He
position is \textit{(h$_0$)} with the solution energy of 5.3-5.8~eV. It was
obtained by moving He from points \textit{h($x=1/2,z=1/4$)}
and \textit{k($x=1/2,z=5/8$)} (Fig.~\ref{FigPositions}a, grey circles). In
\textit{(h$_0$)} He is located in the basal plane near a strongly displaced Be
atom (Fig.~\ref{FigPositions}b).
It is worth to be noted, that this position \textit{(h$_0$)}
was never considered in previous studies 
\cite{Ganchenkova_2009JNuM,Zhang_2012JPCM,Zhang_2012JNM}
as a place for He.

Zero point vibration energy (ZPE) is not small for interstitial He. To
estimate this quantity we calculated phonon frequencies at $\Gamma$-point for
Be$_{96}$He and hcp-Be$_{96}$ supercells within DFPT approach as implemented in
QE software package. ZPE was calculated as a sum of delta functions, that
typically produces 5\% uncertainty of exact value for supercells of such size
(see e.g. ref.~\onlinecite{Sanati_2003SSC}). 
For positions \textit{(b),(d),(h$_0$)} ZPE
contributions to the solution energy are 0.09, 0.08 and 0.15 eV, that does not
change energy landscape significantly. This correlates well with result
by Zhang \textit{et al.} \cite{Zhang_2012JNM}, where ZPE correction was found to
be in narrow range of 0.12-0.13~eV for all studied positions.

A deep potential well for the \textit{(h$_0$)} position, the presence of only
one Be atom in the first coordination shell of He and small He-Be distance
(1.29{\AA}) for this configuration demonstrate the existence of chemical bonding
between He and Be atoms. To get further insight and investigate the nature of
this bonding, we proceeded with the highly-accurate FLAPW method to calculate
the density of electronic states (DOS).
{\squeezetable
\begin{table}[b]
\caption{\label{TabPositions}
Equilibrium positions of He atom in hcp-Be (space group 194),
number of nearest Be atoms $z$, distance He-Be d({\AA}),
calculated by plane-wave pseudopotential 
(PS, QE) and FLAPW (Wien2k) methods for 
Be$_{64}$He, Be$_{96}$He structures.
}
\begin{ruledtabular}
\begin{tabular}{l c c c ccc}
 & Wyckoff & & & \multicolumn{2}{c}{PS} & FLAPW \\ \cline{5-6}
Position & notation & & $z$ & 64 & 96 & 64 \\
 & & & & $d$ & $d$ & $d$ \\ \colrule
(0, 0, 0) & 2 a & unst. & 6 & 1.794 & 1.791 & 1.803 \\
(0, 1/2, 0) & 6 g & unst. & 2 & 1.512 & 1.512 & 1.505 \\
(0, 0, 1/4) & 2 b & stb. & 3 & 1.564 & 1.564 & 1.571 \\
(2/3, 1/3, 1/4) & 2 d & stb. & 3 & 1.726 & 1.719 & 1.740 \\
(0.465, 0.930, 1/4) & 6 h$_0$\footnotemark[1] & stb. & 1 & 1.287 & 1.290 & 1.266 \\
\end{tabular}
\footnotetext[1]{subscript ``0'' indicates minimal solution energy
among all positions}
\end{ruledtabular}
\end{table}
}
{\squeezetable
\begin{table}
\caption{Solution energies of helium (eV) in hcp-Be
from DFT(GGA) calculations.
VASP -- Vienna \textit{Ab initio} Simulation Package 
\cite{VASP_1993PRB,VASP_1996PRB}; ``Y/N'' symbols indicate the presence or absence of
ZPE contribution.}
\label{TabEnergy}
\begin{ruledtabular}
\begin{tabular}{ll c c ccccc}
 Source & Method & ZPE & Cell & \multicolumn{5}{c}{He position} \\
 & & & &  \textit{(a)} & \textit{(g)} & 
 \textit{(b)}& \textit{(d)} & \textit{(h$_0$)} \\\colrule
 &  QE\footnotemark[1] & N & 64  & 6.43  & 6.31 & 6.12 & 6.02 & 5.72 \\
Present & QE & N & 96 & 6.31 & 6.18 & 5.96 & 5.83 & 5.59 \\
study & QE & Y & & unst. & unst. & 6.05 & 5.91 & 5.74 \\
&  Wien2k\footnotemark[1] & N & 64  & 6.06 & 5.91 & 5.72 & 5.78 & 5.34 \\
& & & & & & & \\
Ref.~\onlinecite{Ganchenkova_2009JNuM} & VASP \footnotemark[2] & N & 96 & unst. &
unst. & 5.81 & 5.82  & $\times$ \\
Refs.~\onlinecite{Zhang_2012JPCM},\onlinecite{Zhang_2012JNM} &  VASP \footnotemark[2] &
 Y & 96  & 6.05 & 5.95 & 5.71 & 5.62  & $\times$ \\
\end{tabular}
\footnotetext[1]{With PBE \cite{Perdew_1996PRL} exchange-correlation.}
\footnotetext[2]{With PW91 \cite{Perdew_1992PRB,Burke_1998} exchange-correlation.}
\end{ruledtabular}
\end{table}
}

\begin{figure}[!tp]
\includegraphics[width = 7.5 cm]{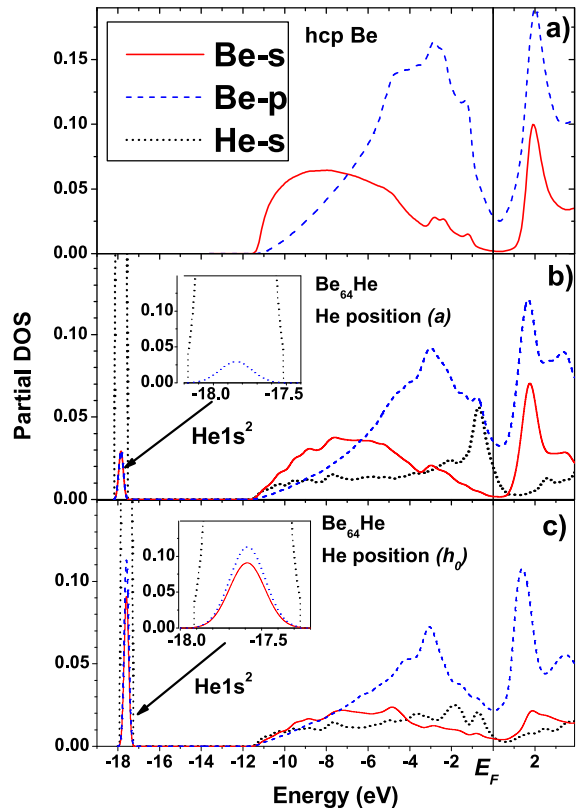}
\caption{(Color online) Partial DOSes of the He-Be system:
a) pure hcp-Be; b) He occupies the octahedral cavity \textit{(a)};
c) He at position \textit{(h$_0$)}.} 
\label{FigDOS}
\end{figure}

Two configurations of He in Be matrix were chosen for comparative DOS analysis,
\textit{(a)} and \textit{(h$_0$)} (Fig.~\ref{FigPositions}).
Figures~\ref{FigDOS}b and ~\ref{FigDOS}c show
partial DOSs of the He-Be system for the cases,
where He occupies the octahedral
cavity \textit{(a)} and the position in the basal plane near the lattice site
\textit{(h$_0$)}. 
Partial DOSs for the pure hcp-Be system are also
shown in Fig.~\ref{FigDOS}a
for comparison. It is seen that the valence band of He-Be system is formed
by the He~$2s$ and the Be $2s$- and $2p$-states. The occupied He $1s$-state is
located 18~eV below the Fermi level.
The densities of electronic states for He in one 
of the \textit{(a)} or
\textit{(h$_0$)} positions appear to be substantially different, which reflects the
different degree of overlap of He electronic charge with that of neighboring Be
atoms.

\begin{figure}[!bt]
\includegraphics[width = 8.5 cm]{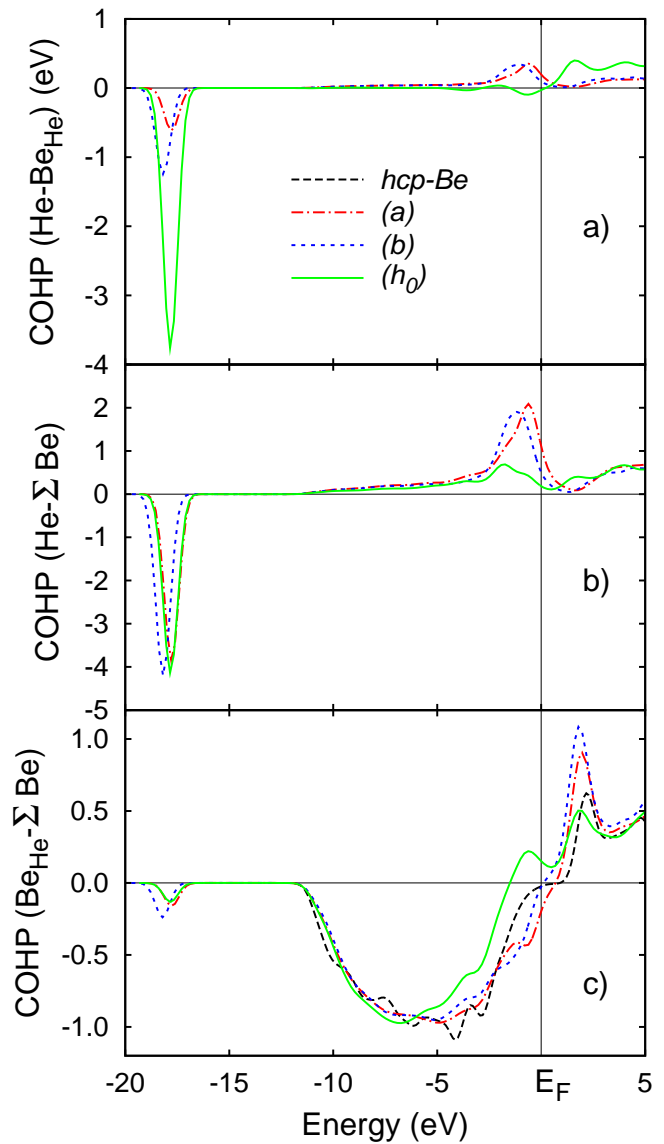}
\caption{(Color online) COHP curves for Be$_{96}$He structure
with helium at positions \textit{(a),(b)} and \textit{(h$_0$)}:
a) He and nearest Be$_\textrm{He}$; 
b) He and beryllium matrix; c) Be$_\textrm{He}$ and beryllium matrix }
\label{FigCOHP}
\end{figure}
Together with substantial decrease of the He-Be distance, the
\textit{(h$_0$)} configuration shows the increase of hybridisation
of the He $s$-states with the $2s$- and $2p$-states of Be. The
hybridisation of low-lying He $1s$-state and $2p$-states of Be also
takes place. We need to mention here that the valence He $s$-states
originate from the empty $2s^0$ atomic level of helium. When the
Be-He system is formed, this empty state moves below the Fermi
level, and becomes partially occupied, forming the $s$-band of Be-He
system. These results reveal an increasing hybridisation of the He
electronic states with those of neighbouring Be atom when helium is
placed in \textit{(h$_0$)} position. Due to the formation of the
chemical Be-He bond, when the He atom is located in the stable
equilibrium \textit{(h$_0$)} position, attraction occurs between He
and the nearest Be atom. As a result, the He and Be atoms are
displaced from the centers of local symmetry and the distance
between them is substantially reduced. The strength of the chemical
bond is nearly equal to the total energy difference between
\textit{(a)} and \textit{(h$_0$)} configurations, $\epsilon_{chem}
\thickapprox \epsilon_a - \epsilon_{h_0} \thickapprox 0.7 eV$.

The analysis of the partial density of states is not enough for complete
understanding of the nature of the chemical bonding between He atom and Be
matrix. We performed additional analysis of the chemical bond formation using
the concept of bonding and antibonding states within the framework of COHP
chemical bond indicator \cite{Dronskowski_1993JPC}. We used pseudopotential
method and local atomic orbital basis set, as realized in SIESTA program package
\cite{Soler_2002JPCM}. The calculations were done in GGA approximation for
exchange-correlation functional \cite{Perdew_1996PRL}, and double-zeta polarized
basis set was used, which included doubled $2s$ and $2p$ orbitals plus
polarizing $3d$ orbital for Be atom, and doubled $1s$ plus polarizing $2p$
orbital for He. 
An optimization of atomic positions of Be$_{96}$He supercell with He in
positions \textit{(a)} or \textit{(h$_0$)} was performed. The obtained relaxed
positions of Be and He atoms were very close to those, obtained by the FLAPW and
pseudopotential (QE) methods. The calculated Be-He bond lengths for positions
\textit{(a),(b),(h$_0$)} were 1.809~{\AA}, 1.580~{\AA}, 1.276~{\AA}, which
agrees well with the results, obtained by other methods
(Table~\ref{TabPositions}).

We plot in Figure~\ref{FigCOHP} the calculated COHP curves for He
and Be atoms for He at positions \textit{(a)},
\textit{(b)} and \textit{(h$_0$)}.
Figure~\ref{FigCOHP}a) shows COHP curves
for He atom and nearest beryllium Be$_\textrm{He}$.
The plot demonstrates that only bonding states are
present for He atom located in \textit{(h$_0$)} position, which is
not the case when He is located in \textit{(a)}
and \textit{(b)}. Antibonding states near Fermi level originate 
from atomic He $2s$-states (see Fig.~\ref{FigDOS}b).
Integrated values (to Fermi level) of these COHP
curves for \textit{(a),(b),(h$_0$)} positions are
0.32, -0.19, -3.42~eV.

COHP curves for He and whole beryllium matrix are shown in Figure~\ref{FigCOHP}b.
The bonding strength of He $1s$-states is the same for all 3 positions, but
reduction of He-Be$_\textrm{He}$ distance (especially strong for
\textit{(h$_0$)}) decreases antibonding states near Fermi level. Integrated COHP
values are 1.93, 1.38, -1.09~eV, confirming He-Be bonding in position
\textit{(h$_0$)}. This fact, as well as the small He-Be$_\textrm{He}$
interatomic distance, allows us to conclude that a distinct molecular 
He-Be$_\textrm{He}$ pair
is formed in Be-hcp matrix.

Soluted in hcp beryllium, helium also changes Be-Be bonding.
Figure~~\ref{FigCOHP}c shows COHP curves for Be$_\textrm{He}$ atom
and beryllium matrix, as well as COHP curve for pure hcp-Be.
It can be seen that the curves for
\textit{(a)}, \textit{(b)} positions are close to that of hcp-Be.
For \textit{(h$_0$)} position, He atom induces antibonding
Be-Be states near Fermi level, thus weakening Be-Be bonding.

\begin{figure}[!tp]
\includegraphics[width = 8.5 cm]{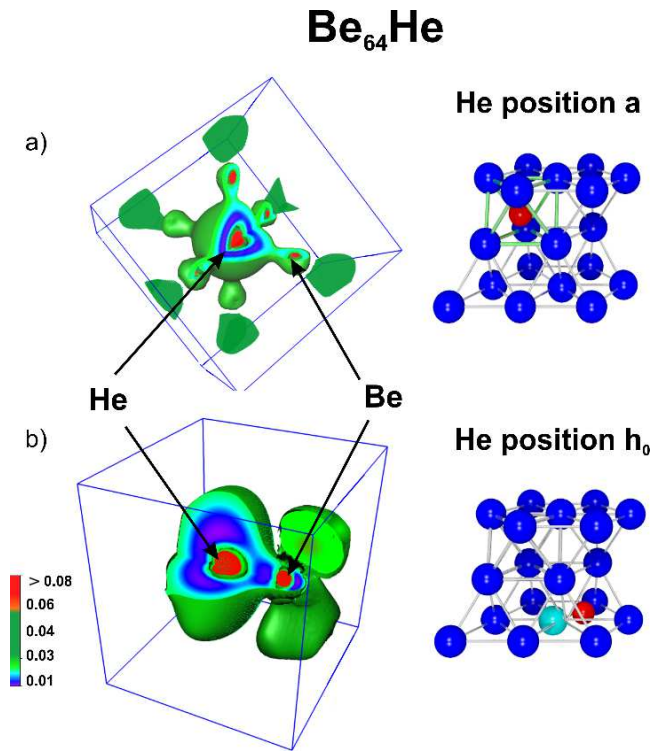}
\caption{(Color online) Distributions of electron density
(e/(a.u.)$^3$) for structure Be$_{64}$He with He atom: a) in
position \textit{(a)} and b) in position \textit{(h$_0$)}. }
\label{FigDensity}
\end{figure}
To visualize the chemical bond in Be-He system, we analyzed the spatial
distribution of valence electronic density in model Be$_{64}$He supercell, when
He is located in \textit{(a)} or \textit{(h$_0$)} positions.
Figure~\ref{FigDensity} shows these distributions for density values
larger than 0.01~e/(a.u.)$^3$. 
When He is located in the octahedral cavity (Fig.\ref{FigDensity}a),
we do not observe electronic charge concentration between He and Be atoms
(marked as red and blue spheres respectively), and the electronic density
(marked as a green cloud) appears to be "pushed" out of the Be octahedron.
Completely different charge distribution is observed in case of He located in
\textit{(h$_0$)} position (Fig.~\ref{FigDensity}b). In this case the electronic
charge "connects" He and Be atoms, which shows a covalent character of chemical
bonding.

In conclusion, our \textit{ab initio} calculations show
a substantial chemical bonding between He and Be atoms, when He
occupies a non-symmetric position in a basal plane in the hcp-Be
matrix. This fact should be taken into account, while analysing the
kinetics of the process of helium accumulation in hcp-Be. The
experimental evidence of the existence of the Be-He bond may be
obtained by means of different methods, including the diffraction
experiments. We believe that our results should stimulate the
experimental efforts in this direction.

\end{document}